\documentclass[12pt]{article}

\usepackage{times}
\usepackage{epsfig}
\usepackage{amssymb}
\usepackage{subfigure}	
\usepackage{graphicx}	
\usepackage{color}					\usepackage{jheppub}

\begin{document}

\newcommand{\nc}{\newcommand}
\newcommand{\rnc}{\renewcommand}

\rnc{\baselinestretch}{1.24}    
\setlength{\jot}{6pt}       
\rnc{\arraystretch}{1.24}   

\makeatletter
\rnc{\theequation}{\thesection.\arabic{equation}}
\@addtoreset{equation}{section}
\makeatother


 \renewcommand{\thefootnote}{\fnsymbol{footnote}}

\newcommand{\tcb}{\textcolor{blue}}
\newcommand{\tcr}{\textcolor{red}}
\newcommand{\tcg}{\textcolor{green}}


\def\be{\begin{eqnarray}}
\def\ee{\end{eqnarray}}
\def\nn{\nonumber\\}


\def\ct{\cite}
\def\la{\label}
\def\eq#1{\eqref{#1}}


\def\a{\alpha}
\def\b{\beta}
\def\g{\gamma}
\def\G{\Gamma}
\def\d{\delta}
\def\D{\Delta}
\def\e{\epsilon}
\def\et{\eta}
\def\ph{\phi}
\def\Ph{\Phi}
\def\ps{\psi}
\def\Ps{\Psi}
\def\k{\kappa}
\def\l{\lambda}
\def\L{\Lambda}
\def\m{\mu}
\def\n{\nu}
\def\th{\theta}
\def\Th{\Theta}
\def\r{\rho}
\def\s{\sigma}
\def\S{\Sigma}
\def\ta{\tau}
\def\o{\omega}
\def\O{\Omega}
\def\pr{\prime}


\def\half{\frac{1}{2}}
\def\goto{\rightarrow}

\def\na{\nabla}
\def\grad{\nabla}
\def\curl{\nabla\times}
\def\div{\nabla\cdot}
\def\pa{\partial}
\def\fr{\frac}

\def\bra{\left\langle}
\def\ket{\right\rangle}
\def\lb{\left[}
\def\lc{\left\{}
\def\ls{\left(}
\def\lp{\left.}
\def\rp{\right.}
\def\rb{\right]}
\def\rc{\right\}}
\def\rs{\right)}

\def\vac#1{\mid #1 \rangle}


\def\td#1{\tilde{#1}}
\def\check{ \maltese {\bf Check!}}


\def\Tr{{\rm Tr}\,}
\def\det{{\rm det}}
\def\text#1{{\rm #1}}


\def\bc#1{\nnindent {\bf $\bullet$ #1} \\ }
\def\ch {$<Check!>$ }
\def\ss {\vspace{1.5cm}}
\def\inf{\infty}

\begin{titlepage}

\hfill\parbox{2cm} { }

%
\vspace{1cm}

\begin{center}
{\Large \bf Holographic description for correlation functions}

\vskip 1. cm
 {Hanse Kim$^{a}$\footnote{e-mail : powerblo@gist.ac.kr}},
 {Jitendra Pal$^{b}$\footnote{e-mail : jpal1@ph.iitr.ac.in}}, and
 {Chanyong Park$^{a}$\footnote{e-mail : cyong21@gist.ac.kr}}

\vskip 1cm

{\it $^a$ Department of Physics and Photon Science, Gwangju Institute of Science and Technology \\ 
 Gwangju  61005, Korea}
 
 {\it $^b$ Department of Physics, Indian Institute of Technology Roorkee\\Roorkee
247667 Uttarakhand, India  }

\end{center}

\thispagestyle{empty}

\vskip2cm


\centerline{\bf ABSTRACT} 

\vspace{1cm}

We study general correlation functions of various quantum field theories in the holographic setup. Following the holographic proposal, we investigate correlation functions via a geodesic length connecting boundary operators. We show that this holographic description can reproduce the known two- and three-point functions of conformal field theory. Using this holographic method, we further study general two-point functions of a two-dimensional thermal CFT and of a scalar field theory living in a dS or AdS space. Due to the nontrivial thermal or curvature effect, the two-point functions in an IR limit show different scaling behaviors from those of the UV CFT. We study such nontrivial IR scaling behaviors by applying the holographic method.


\vspace{2cm}

\end{titlepage}

\renewcommand{\thefootnote}{\arabic{footnote}}
\setcounter{footnote}{0}



\section{Introduction}

After the AdS/CFT correspondence proposal, there were many attempts to account for 
strongly interacting systems on the holographic dual gravity side. The AdS/CFT conjecture allows us to relate a $(d+1)$- dimensional classical gravity theory to a $d$-dimensional non-gravitational conformal field theory (CFT) \cite{Maldacena:1997re, Gubser:1998bc, Witten:1998qj, Witten:1998zw}. Moreover, the AdS/CFT correspondence proposed that the gravity theory can give us information about the nonperturbative features of a quantum field theory (QFT). In this case, the extra dimension in the bulk is identified with the energy scale observing the dual QFT. Therefore, the gravity theory maps to the renormalization group (RG) flow of the dual QFT. From the RG flow point of view, non-abelian gauge theories are weakly interacting at a UV fixed point and have a conformal symmetry. In an IR limit, on the other hand, they are strongly interacting and reveal various non-perturbative features. In the situation without a well-established mathematical method describing the non-perturbative RG flow, the AdS/CFT correspondence can provide a new method to look into the non-perturbative IR physics. In this work, we investigate general correlation functions of QFT on the dual gravity theory side, with an emphasis on studying their scale dependence.

The main focus of this work is the study of the correlation functions of various QFT at finite temperature or in the curved spacetime. In general, computing non-perturbative correlation functions on the QFT side is a difficult task because all loop quantum corrections must be taken into account. This motivates us to look at the holographic setup in order to calculate non-perturbative correlation functions. One of the interesting features of the AdS/CFT correspondence is that many important physical quantities, like $q\bar{q}$ potential\cite{Maldacena:1998im,Rey:1998bq,Park:2009nb}  and entanglement entropy\cite{Ryu:2006bv,Ryu:2006ef,Myers:2012ed,Blanco:2013joa,Wong:2013gua,Bhattacharya:2012mi,Momeni:2015vka,Fischler:2012ca,Kim:2016jwu,Park:2015dia,Park:2021wep}, can be realized as geometrical objects on the dual gravity side. Similarly, it was also conjectured that a two-point function maps to a geodesic curve connecting two boundary operators  
\be			\la{Formula:twopoint}
\bra O (\ta_1,\vec{x}_1) \ O(\ta_2, \vec{x}_2) \ket = e^{- \D \,  L (\ta_1, \vec{x}_1;\ta_2,\vec{x}_2) /R} ,
\ee
where $\D$ is the conformal dimension of a local operator $O (\ta,\vec{x}) $ and $ L (\ta_1, \vec{x}_1;\ta_2,\vec{x}_2)$ indicates a geodesic length connecting two local operators.

Quantum correlation plays a crucial role in understanding physics both in the weak and strong coupling limits. Despite its importance, it is not easy to calculate correlation functions nonpertubatively. Although the perturbative method are applicable to UV theories, it is not valid in the IR region having a strong coupling constant. The IR physics reveals many new nontrivial physical phenomena which cannot be explained by the perturbative methods. One way to understand such nontrivial IR physics is to take into account the non-perturbative RG flow. The AdS/CFT correspondence may shed lights on constructing the non-perturbative RG flow and understanding the IR physics correctly. Therefore, it would be interesting to investigate how to evaluate the correlation functions in the holographic dual gravity. In this work, we study how to reproduce the most general two- and three-point correlation functions in the dual gravity. We first consider a Euclidean AdS space and then evaluate the Euclidean correlation functions by considering a geodesic curve connecting two boundary operators. After the Wick rotation, we rederived the known Lorentzian correlation functions. This holographic method is further applied to the BTZ black hole, which is the dual of a two-dimensional thermal CFT, and to an AdS space with a dS boundary, which is the dual of a CFT living in an eternally inflating universe.

Computing correlation function of thermal systems \cite{ Rod1, Rod2,Fuertes,Geo1, Park:2022abi, Park:2022mxj,Krishna,Papadodimas} is one of interesting area of research. For thermal systems, although the finite thermal corrections are negligible in the UV  region, they can give rise to a significant effect on the IR physics. Due to this, the IR physics can show a new physics law like a thermodynamic relation which cannot be explained by the fundamental UV theory. In order to understand such new macroscopic orders, it would be important to know how the correlations of operators depend on the energy scale. By applying the holographic technique, we calculate the most general two-point function of a thermal two-dimensional CFT \cite{Kraus,Shenker,Hamilton,Balasubramanian,Polchinski,Papadodimas1,Hartman,Ooguri,Grinberg}. We show that, although a thermal CFT is conformal at the UV fixed point, the screening effect caused by the background thermal fluctuations provides an effective mass to local operators and then results in the exponential suppression of the two-point function. In this case, the effective mass is proportional to temperature.

We also investigate the correlation functions in the curved spacetime like an eternally expanding universe. The dS space describing an eternal inflation is one of the important background geometry to understand the birth and evolution of our universe. The correlation functions on that background also provide several important information, like a power spectrum and non-Gaussianity, to understand the history of our universe. In the cosmology studies, there were many studies about the correlation function of massive and massless scalar fields on the de Sitter background \cite{Thirring,Chernikov,Tagirov,Bunch,Linde,Starobinsky, Vilenkin, Allen1, Allen2,Schlingemann:1999mk}. In this work, we look into the correlation function of a dS space by applying the holographic method. To do so, we take into account an $(d+1)$-dimensional AdS space whose boundary is given by a $d$-dimensional dS space. After calculating the geodesic length connecting two operators living in the dS boundary, we derive the most general two-point function in the eternally expanding universe. We also show that this holographic result is coincident with the result of the fress scalar field theory defined in a $d$-dimensional dS space. This work is further generalized to the correlation function in a $d$-dimensional AdS space.


\section{Holographic description for correlation functions }


For CFTs with a Lorentzian signature, the conformal symmetry uniquely determines two- and three-point correlation functions up to an overall constant. Denoting the distance between two operators as $|r_1 - r_2| = \sqrt{- |t_1 - t_2|^2 + | \vec{x}_1 - \vec{x}_2|^2}$, two- and three-point functions are given by
\be
\bra O (t_1 , \vec{r}_1) \ O (t_2 , \vec{r}_2) \ket &=&  \fr{N}{ | r_1 - r_2 |^{ 2 \D} } ,  \la{Result:CFT2pt} \\
\bra O(t_1,r_1) \ O(t_2, r_2) \ O(t_3, r_3) \ket &=& \fr{C_{123}}{ |r_1 - r_2|^{\D_1 +\D_2 -\D_3 } \  |r_2 - r_3|^{\D_2 +\D_3 -\D_1 } \ |r_3 - r_1|^{\D_3 +\D_1 -\D_2 } } ,
\ee 
where $\D = \D_1 =\D_2$ and $\D_i$ means a conformal dimension of an $i$-th operator, $O(t_i,\vec{x}_i)$. Normalizaing operators, without loss of generality, allows us to set $N=1$. The structure constant $C_{123}$ for the three-point function corresponds to the operator product expansion coefficient which cannot be fixed by the conformal symmetry. According to the AdS/CFT correspondence, a $d$-dimensional CFT has a one-to-one map to a gravity defined on a $(d+1)$-dimensional AdS space. Therefore, it would be interesting to reproduce the above general correlators in the holographic setup. This holographic study on correlators may be helpful to understand the microscopic and macroscopic correlation for interacting QFTs nonpertubatively.

For comparison with correlators of CFT at finite temperature or in curved spaces, we first briefly discuss how to calculate a two-point function in a $d$-dimensional Euclidean space having a $SO(d)$ rotational symmetry, A metric of a $d$-dimensional Euclidean space is given by
\be
ds^2 = \d_{\m\n} dx^\m dx^\n = d \ta^2 + d \vec{x} \cdot d \vec{x} .
\ee
On this flat background geometry, we take into account a massless scalar field theory 
\be
S = \half \int d^d x \,  \ls \pa \ph \rs^2 .
\ee
The free scalar field theory is conformal and the conformal dimension of $\ph$ is given by  $\D_\ph = (d-2)/2$. In this case, a two-point function of $\ph$ is determined by the following equation
\be
- \, \pa^\m   \pa_\m  \bra \ph (\ta_1,\vec{x}_1) \ \ph (\ta_2,\vec{x}_2) \ket  = \d^{(d)} (|r_1-r_2|) ,
\ee
where $|r_1 - r_2|=\sqrt{ | \ta_1- \ta_2 |^2 +  | \vec{x}_1- \vec{x}_2 |^2 }$. A general solution satisfying this equation is given by
\be
\bra \ph (\ta_1,\vec{x}_1 ) \ \ph (\ta_2,\vec{x}_2) \ket \sim  \fr{1}{ | r_1- r_2 |^{ 2 \D_\ph} } .
\ee
 If we further consider an operator, $O = \ph^n$, the Wick contraction allows a general two-point function to be 
\be
\bra O (\ta_1 , \vec{x}_1) \ O (\ta_2 , \vec{x}_2) \ket \sim  \fr{1}{ \ls   |\ta_1- \ta_2 |^2 +  | \vec{x}_1- \vec{x}_2 |^2  \rs^{  \D}  } ,   \la{Result:2funCFT} 
\ee
where the conformal dimension of $O$ is given by $\D = n \D_\ph$.

After the Wick rotation ($\ta = i t$), the Euclidean two-point function reduces to a Lorentzian one
\be
\bra O (t_1 , \vec{x}_1) \ O (t_2 , \vec{x}_2) \ket \sim  \fr{1}{  \ls  - | t_1- t_2 |^2 +  | \vec{x}_1- \vec{x}_2 |^2   \rs^{  \D} } ,  \la{Result:L2funCFT} 
\ee
where $\sqrt{ - | t_1- t_2 |^2 +  | \vec{x}_1- \vec{x}_2 |^2 }$ means a proper distance preserving the $SO(1,d-1)$ Lorentz symmetry. This is the two-point function expected by the conformal symmetry. According to the AdS/CFT correspondence, it must be possible to reproduce this general two-point functions in the dual AdS gravity. In the previous works \cite{Park:2022abi,Park:2022mxj}, a spatial (equal-time) correlator at $\ta_1=\ta_2$ and a temporal (equal-position or autocorrelation) two-point function at $\vec{x}_1 = \vec{x}_2$ have been studied. In this work, we further investigate how to calculate the general two-point functions for $\ta_1\ne \ta_2$ and $\vec{x}_1\ne \vec{x}_2$ in the holographic setup.



In the holographic study, there are two different ways to obtain a two-point function. One is to consider a bulk field which is dual of a local operator of the boundary theory. After calculating a bulk-to-boundary Green function, we move the bulk field to the boundary to obtain a boundary-to-boundary Green function which is identified with a two-point function of the dual field theory. It was shown that this boundary-to-boundary Green function leads to the above two-point function in \eq{Result:CFT2pt}. Due to the direct relation between the bulk field and its dual operator, this method is useful to understand more internal structure of a two-point function. However, if we consider a non-AdS geometry, it is usually hard to find  bulk-to-boundary Green function. In this case, there is another way to evaluate a two-point function. It was proposed that a two-point function can be described by a geodesic length connecting two boundary operators
\be
\bra O(\ta_1,\vec{x}_1) \ O(\ta_2, \vec{x}_2) \ket = e^{- \D \,  L (\ta_1, \vec{x}_1;\ta_2,\vec{x}_2) /R} ,   \la{Formula:TPF}
\ee
where $\D$ is the conformal dimension of the boundary operator. Using this proposal, we can reproduce the previous CFT's general two- and three-point function exactly. Furthermore, we can also calculate general two-point functions of CFTs at finite temperature or in curved spaces, as will be seen later. 

We discuss how to calculate the previous general two- and three-point functions holographically. For convenience, we first calculate Euclidean correlation functions and then obtain Lorentzian ones by applying the Wick rotation. In order to describe a $d$-dimensional CFT holographically, we take into account a $(d+1)$-dimensional Euclidean AdS space  whose metric is given by
\be
ds^2 = \fr{R^2}{z^2} \ls d \ta^2 + d \vec{x} \cdot d \vec{x} + dz^2 \rs ,
\ee
where the dual CFT lives at the boundary ($z=0$).

In order to evaluate general two- and three-point functions, we first investigate a bulk-to-boundary Green function running from $\lc \ta, \vec{x} , z \rc = \lc \ta_J, \vec{x}_J , z_J \rc$ to $\lc \ta_i, \vec{x}_i, 0 \rc$ which corresponds to the position of an $i$-th operator $O(\ta_i, \vec{x}_i)$. To find a geodesic connecting these two points, we regard $\vec{x}$ and $z$ as functions of $\ta$. Then, a geodesic length is governed by
\be
L (\ta_J, \vec{x}_J,z_J;\ta_1,\vec{x}_1, 0) \equiv  \int_{\ta_1}^{ \ta_J}  d\ta {\cal L} = R \int_{\ta_1}^{\ta_J}  d \ta \ \fr{ \sqrt{ 1  + \dot{x}^2 + \dot{z}^2} }{z}  ,  \la{Action:dualCFT}
\ee
where $x = | \vec{x}|$ and the dot means a derivative with respect to $\ta$. Since the geodesic depends on $\ta$ and $x$ implicitly, there exist two conserved charges. The first one is the canonical momentum of $\vec{x}$ 
\be
\vec{P} = \fr{\pa {\cal L} }{\pa \dot{\vec{x}}}  = \fr{R \, \dot{\vec{x}}}{z \sqrt{ 1  + \dot{x}^2 + \dot{z}^2}} ,  \la{Result:consevedP}
\ee
and the other is Hamiltonian corresponding to the canonical momentum of $\ta$ 
\be
H = \fr{\pa {\cal L} }{\pa \dot{\vec{x}}} \ \dot{\vec{x}} + \fr{\pa {\cal L} }{\pa \dot{z}} \ \dot{z} - {\cal L} = - \fr{R}{z \sqrt{ 1  + \dot{x}^2 + \dot{z}^2}}  . \la{Result:consevedH}
\ee
To fix the conserved charges, we introduce a turning point  $\lc \ta_t, r_t, z_t \rc$ where $\dot{z} (\ta_t) = 0$.  Then, the conserved charges at a turning point reduce to
\be
\vec{P} &=& \fr{R \, \vec{v}}{z_t \sqrt{ 1  + v^2 }}  ,  \la{Result:conP0} \\
H&=& - \fr{R}{z_t \sqrt{ 1  + v^2 }} ,  \la{Equation:radial}
\ee
where $v = |\vec{v}|$ denotes a velocity at a turning point, $\vec{v} = \dot{\vec{x}} (\ta_t)$.

Comparing these conserved charges, we find that $ \dot{r}$ is given by a constant 
\be
\fr{d \vec{x}}{d \ta} = \vec{v} .   \la{Equation:radialr}
\ee 
A solution satisfying two boundary conditions, $x_1 = x(\ta_1)$ and $x_J = x (\ta_J)$, reduces to 
\be
\vec{x} (\ta) = \vec{v} \, ( \ta - \ta_1 ) + \vec{x}_1  .
\ee
where $\vec{v} = \D \vec{x}/\D \ta$ with $\D \vec{x}=  \vec{x}_J - \vec{x}_1$ and $\D \ta = \ta_J-\ta_1 $. Moreover, comparing \eq{Result:consevedH} with \eq{Equation:radial} leads to an equation governing the radial motion of a geodesic
\be
\fr{dz}{d\ta} =  \pm \fr{  \sqrt{1+v^2} \, \sqrt{z_t^2 - z^2}}{z} .  \la{Equation:radialz}
\ee
A solution satisfying $0 = z(\ta_1)$ and $z_J = z (\ta_J)$ becomes
\be
z (\ta) =  \sqrt{  \frac{ ( \D \ta^2 + \D x^2  +z_J^2 ) \,  \left( \tau -\tau _1 \right) }{\D \ta  }  - \fr{\left(\D \ta^2 + \D x^2  \right) \, \left(\tau -\tau _1\right)^2}{\D \ta^2}  }
\ee
where a turning point appears at
\be
z_t = \frac{\D \ta^2+ \D x^2+ z_J^2}{2 \sqrt{\D \ta^2 +  \D x^2}} .
\ee

After substituting \eq{Equation:radialr} and \eq{Equation:radialz} into \eq{Action:dualCFT}, the integration of \eq{Action:dualCFT} for $ \ta_1< \ta_t <\ta_J$  results in
\be
L (\ta_J,x_J,z_J;\ta_1,x_1, 0) &=& R \ \ls  \int_\e^{z_t} dz  -  \int_{z_t}^{z_J} dz \rs \ 
\fr{ z_t}{z \sqrt{z_t^2 - z^2}}   \nn
&=&  R \log \fr{\D \ta^2+ \D x^2+ z_J^2 }{z_J} - R \log \e  ,   \la{Result:Junctiontoboundary}
\ee
where $\e$ is introduced as a UV cutoff. Following \eq{Formula:TPF},  a bulk-to-boundary Green function reads
\be
\bra  O (\ta_J,\vec{x}_J,z_J) \ O (\ta_1,\vec{x}_1, \e)  \ket = \fr{z_J^{\D} \  \e^\D}{\ls\D \ta^2+ \D x^2+ z_J^2  \rs^\D} .
\ee
If we take $z_J \to \e$, the bulk-to-boundary Green function further reduces to a boundary-to-boundary Green function or a general two-point function of dual CFT
\be
\bra  {\cal O} (\ta_J,\vec{x}_J) \ {\cal O} (\ta_1,\vec{x}_1)  \ket = \fr{ 1 }{\ls  \D \ta^2+ \D x^2 \rs^{  \D}} ,
\ee
where an normalized operator is defined as ${\cal O} \equiv O /\e^\D$. After the Wick rotation, we finally obtain the Lorentzian two-point function in \eq{Result:L2funCFT}


Now, we take into account the correlation function of three operators located at arbitrary positions. In Ref. \ct{Rod1,Keranen:2016ija}, it was studied how to evaluate a three-point autocorrelation function when three operators are located at the same spatial position. In order to describe a three-point function holographically, we introduce a junction point in the bulk, whose position is denoted by $\lc \ta, \vec{x}, z \rc = \lc \ta_J,\vec{x}_J, z_J \rc$. Then, a three-point function can be determined as the sum of minimal geodesic lengths connecting the junction point to three boundary operators
\be
\bra O(\ta_1,\vec{x}_1) \ O(\ta_2, \vec{x}_2) \ O(\ta_3, \vec{x}_3) \ket = \exp \ls {- \fr{\sum_{i=1}^{3} \D_i\,  L_i (\ta_i, \vec{x}_i,0; \ta_J, \vec{x}_J, z_J) }{R}} \rs ,   \la{Formula:threeptfun}
\ee
where $\D_i$ indicates the conformal dimension of an $i$-th boundary operator.
Using the previous bulk-to-boundary Green function in \eq{Result:Junctiontoboundary}, a geodesic length connecting three operators via the junction point is given by  
\be
\fr{\sum_{i=1}^{3} \D_i \,  L_i (\ta_i, \vec{x}_i,0; \ta_J,\vec{x}_J, z_J)}{ R }
&=& \sum_{i=1}^{3} \D_i \ls \log \fr{ \D \ta_i^2+ \D x_i^2+ z_J^2 }{z_J} - \log \e \rs .
\ee
In this case, the junction point is determined as a position which minimize the the geodesic length. After variation, the junction point must satisfy
\be
0 &=& \sum_{i=1}^{3} \fr{\D_i \  \D r_i}{ \D r_i^2 + z_J^2} , \nn
0 &=& \sum_{i=1}^{3} \fr{ \D_i \ (\D r_i^2  - z_J^2) }{ \D r_i^2 + z_J^2} .
\ee
Solving these equations determines the position of the junction point. To do so, we parameterize $\D \ta_i$ and $\D x_i $ as the following form 
\be
\D \ta_i = \D r_i  \cos \D \th_i \quad {\rm and} \quad \D x_i = \D r_i \sin \D\th_i ,
\ee
where $\D r_i = |r_J - r_i | =\sqrt{\D \ta_i^2 + \D x_i^2 }$ with $\D x_i=  |\vec{x}_J - \vec{x}_i|$, $\D \ta_i = | \ta_J-\ta_i|$, and $\D\th_i= | \th_J - \th_i | $. Since the geodesic length is invariant under the rotation in the $\ta-x$ plane, we can set $\D \ta_i = \D r_i $ with $\D \th_i = 0$ without loss of generality. This implies that a general three-point function  up to the rotation in the $\ta-x$ plane is the same as the three-point autocorrelation function studied in Ref \ct{Keranen:2016ija}. As a result, a general three-point function is given by
\be
\bra O(\ta_1,\vec{x}_1) \ O(\ta_2, \vec{x}_2) \ O(\ta_3, \vec{x}_3) \ket = \fr{C_{123}}{ |r_1 - r_2|^{\D_1 +\D_2 -\D_3 } \  |r_2 - r_3|^{\D_2 +\D_3 -\D_1 } \ |r_3 - r_1|^{\D_3 +\D_1 -\D_2 } } ,
\ee
where $|r_i-r_j| = \sqrt{|\ta_i-\ta_j|^2 + |x_i-x_j|^2}$ and a structure constant $C_{123}$ is given by
\be
C_{123} =   \fr{ \gamma^{\Delta } }{2^{\Delta} \Delta _1^{\Delta _1} \Delta _2^{\Delta _2} 
   \Delta _3^{\Delta _3} \left(\Delta _1-\Delta _2-\Delta
   _3\right){}^{\Delta _1} \left(\Delta _2-\Delta _3-\Delta _1 \right){}^{\Delta _2} \left(\Delta _3 -\Delta _1 -\Delta _2\right){}^{\Delta _3}  } ,
\ee
with
\be
\D &=& \D_1 + \D_2 + \D_3 , \nn
\g &=& 2 \ls \Delta _2^1 \Delta _2^2+  \Delta _3^2 \Delta _1^2 +  \Delta _2^2 \Delta _3^2 \rs  
-\Delta _1^4 -\Delta _2^4-\Delta _3^4 .
\ee 
This is consistent with the three-point function expected in CFT. 

The holographic method studied here is also applicable to other theories. For example, we can regard a thermal CFT at finite temperature. Although a finite temperature effect is negligible in the UV region, it in the IR region crucially modifies the IR correlators. To study this thermal effect, we calculate a two-point function of a two-dimensional thermal CFT whose dual gravity is described by a BTZ black hole 
\be
ds^2 = \fr{R^2}{z^2} \ls f(z) d \ta^2 + d x^2  + \fr{1}{f(z)} \ dz^2 \rs 
\ee
with a blackening factor
\be
f(z) = 1 - \fr{z^2}{z_h^2} .
\ee
For $z_h \to \infty$, the black hole geometry reduces to an AdS space corresponding to the zero temperature limit. To apply \eq{Formula:TPF}, we first calculate a geodesic length extending to the BTZ black hole geometry. First, we assume that two local operators are located at the boundary, $\lc z,\ta, x  \rc= \lc 0, \ta_1, x_1 \rc$ and $\lc  0, \ta_2, x_2   \rc$. Due to the translational symmetry in $\ta$ and $x$ directions, the two-point function is reexpressed as the following form 
\be
 \bra O(\ta_1,x_1) \ O(\ta_2, x_2) \ket =  \bra O \ls -\D \ta /2, - \D x /2 \rs \ O \ls \D \ta /2,  \D x /2 \rs \ket ,
\ee
where $\D \ta = |\ta_1 - \ta_2|$ and $\D x = |x_1 - x_2|$. Then, the geodesic length in the black hole geometry is governed by  
\be
L (\ta_1,x_1;\ta_2,x_2)   = R \int_{- \D \ta/2}^{\D \ta /2}  d \ta \ \fr{ \sqrt{ f^2  + f \dot{x}^2 + \dot{z}^2} }{z \, \sqrt{f}} .
\ee

The geodesic length, similar to the previous case, relies on $\ta$ and $x$ implicitly, so that the canonical momenta of $x$ and $\ta$ are conserved. More precisely, the canonical momenta of $x$ and $\ta$ are given by
\be
P 
= \fr{ R \, \dot{x} \sqrt{f}}{ z \sqrt{ f^2  + f \dot{x}^2 + \dot{z}^2} } ,  \la{Result:Con1} \quad {\rm and} \quad
H 
= \fr{R  \, f^{3/2}}{ z \sqrt{ f^2  + f \dot{x}^2 + \dot{z}^2} }  .
\ee
In this case, a turning point appears at $\lc  \ta, x, z \rc = \lc  0, 0, z_t \rc$ due to the invariance under $\ta \to - \ta$ and $x \to -x$. At the turning point, the conserved quantities read
\be
P  =  \fr{ R \, v }{ z_t \sqrt{ f_t  +  v^2 } }  ,   \quad {\rm and} \quad H  =  \fr{ R \, f_t }{ z_t \sqrt{ f_t  + v^2 } }  ,
\ee
where $f_t$ and $v$ are the values of $f$ and $\dot{x}$ at the turning point. Comparing the conserved quantities, we find the relations between $x$, $z$ and $\ta$  
\be
\fr{dx}{d \ta} &=& \fr{f v}{f_t}  \la{Relation:v}  , \la{Relation:dxdt} \\
\fr{dz}{d \ta} &=& \frac{\left(z_h^2-z^2\right) \sqrt{z_t^2-z^2}  \sqrt{\left(v^2+1\right) z_h^2-v^2 z^2-z_t^2}}{z_h  \left(z_h^2-z_t^2\right) z } .  \la{Relation:dzdt}
\ee 
Solving \eq{Relation:dzdt}, we can find $z_t$ and $v$ as functions of $\D \ta$ and $\D x$
\be
z_t &=& z_h \sin \ls \fr{\D \ta}{2 z_h} \rs  \sqrt{ 1+ \cot \ls \fr{ \D \ta}{2 z_h} \rs^2 \tanh \ls \fr{\D x}{2 z_h} \rs ^2} , \\
v &=& \cot \ls \fr{\D \ta}{2 z_h} \rs \tanh \ls \fr{\D x}{2 z_h} \rs   .
\ee
These results finally determine the geodesic length as the following form
\be
L (\ta_1, x_1;\ta_2, x_2) = R \lb 2 \log \fr{1}{\e}  + \log \lc 2 z_h^2 \cosh  \ls \fr{\D x}{z_h} \rs -  2 z_h^2 \cos \ls \fr{\D \ta}{z_h} \rs \rc \rb .
\ee

According to the holographic proposal in \eq{Formula:TPF}, a general Euclidean two-point function  becomes up to  normalization  
\be
\bra O \ls \ta_1, x_1\rs \ O \ls \ta_2,  x_2 \rs \ket  
\sim \fr{1}{ \left|    \sin^2 \ls \fr{|\ta_1 - \ta_2|}{ 2 z_h} \rs  + \sinh^2  \ls \fr{| x_1 - x_2|}{2 z_h} \rs   \right|^{\D} }   .   \la{Result:Eth2pt}
\ee
In a UV limit with a short distance and time interval ($\D \ta, \D x \to 0$), the thermal two-point function reduces to the previous CFT's result in \eq{Result:2funCFT}. This is because finite thermal corrections are negligible in the UV region. Applying the Wick rotation to \eq{Result:Eth2pt}, a general Lorentzian two-point function is rewritten as  \cite{Rod1}
\be
\bra O \ls t_1, x_1\rs \ O \ls t_2,  x_2 \rs \ket   
&\sim&  \fr{1}{ \left|  -   \sinh^2 \ls \fr{|t_1 - t_2|}{2z_h} \rs  + \sinh^2  \ls \fr{ | x_1 - x_2| }{2 z_h} \rs \right|^{\D} } . \la{Result:Lth2pt}
\ee

In the IR region having a long distance ($| x_1 - x_2| \gg  |t_1 - t_2| \gg z_h$), a spatial two-point function reduces to
\be
\bra O ( t_1, x_1) O( t_2, x_2) \ket  \sim e^{-   | x_1-x_2|  /\xi}  ,
\ee
with a correlation length $\xi$ or the inverse of an effective mass $m_{eff}$ 
\be
\xi \equiv \fr{1}{m_{eff}  } = \fr{1}{2  \pi  \D   T_H}   ,
\ee
where $T_H = 1 /(2\pi z_h)$ means the Hawking temperature. Exponential suppression of the thermal correlator in the IR limit is due to the screening effect of thermal background. In another IR limit with a long time interval ($|t_1 - t_2| \gg | x_1 - x_2|  \gg z_h$), a similar feature also appears for a temporal two-point function  
\be
\bra O ( t_1, x_1) O( t_2, x_2) \ket  \sim  e^{-  |t_1-t_2|  / \, t_{1/2} } ,
\ee
where a half-life time $t_{1/2}$ is given by the inverse of the effective mass, $t_{1/2} = 1/m_{eff}$. As expected before, the IR correlators in the thermal CFT behave totally different from the UV ones. In the UV region, the correlation suppresses by a power law due to the conformal symmetry, while the IR correlation exponentially suppresses due to the screening effect of the thermal background.

\section{Correlation functions in curved spaces}

Now, we consider a holographic dual of QFTs living in curved spaces like a dS or AdS space and then investigate correlation functions of such theories. On the QFT side, two-point correlators of a massive field in a global dS space was studied in Ref. \cite{Aalsma:2022eru,Chapman:2022mqd}. In order to describe QFTs living in a $d$-dimensional dS and AdS space holographically, we have to take into account a $(d+1)$-dimensional AdS space whose boundary is given by a dS or AdS space. To do so, let us first  consider a $(d+1)$-dimensional Poincare AdS space
\be
ds^2 = \fr{R^2}{z^2}  \ls - dt^2 + \d_{ij} \, dx^i dx^j + dz^2 \rs ,   \la{Metric:d1AdS}
\ee
where $i,j = 1 \cdots d-1$. This AdS metric has a $d$-dimensional flat boundary at $z=0$. To describe an AdS space with an AdS boundary, we introduce new coordinates, $u$ and $y$,
\be 
z = \fr{u \, w}{R} \ , \quad  x^a = y^a  \ {\rm and } \quad   x^{d-1} =  w \sqrt{1 - \fr{u^2}{R^2} }  ,      \la{Relation:coord}
\ee
where $a,b = 1, \cdots, d-2$. Then, the AdS metric is rewritten as
\be
ds^2 = \fr{R^2}{u^2} \lb \fr{du^2}{1 - u^2/R^2}  + \fr{R^2}{w^2} \ls  - d t^2 + \d_{ab} \, dy^a dy^b + dw^2\rs\rb .
\ee
At $u=\e$ in the limit of $\e \to 0$, the boundary reduces to a $d$-dimensional AdS space
\be
ds_{AdS}^2 =   \fr{\td{R}^2}{w^2} \ls  - d t^2 + \d_{ab} \,  dy^a dy^b + dw^2  \rs  ,
\ee
where $\td{R} = R^2/\e$. Applying the Wick rotation ($\ta = i t$), the Euclidean bulk metric becomes 
\be
ds^2 = \fr{R^2}{u^2} \lb \fr{du^2}{1 - u^2/R^2}  + \fr{R^2}{w^2} \ls  d \ta^2 + \d_{ab} \, dy^a dy^b + dw^2 \rs\rb .   \la{Result:EAdSwithAdS}
\ee
This is the metric of a  $(d+1)$-dimensional Euclidean AdS space with a $d$-dimensional AdS boundary.

We can also consider an AdS space having a dS boundary. Introducing another coordinate system
\be
z = \fr{T \bar{u}}{R} \quad {\rm and} \quad t= T \, \sqrt{1 + \fr{\bar{u}^2}{R^2}} ,
\ee
the bulk AdS metric \eq{Metric:d1AdS} is rewritten as
\be		 
ds^2 = \fr{R^2}{\bar{u}^2} \lb \fr{d \bar{u}^2}{1+\bar{u}^2/R^2}  + \fr{R^2}{T^2} \ls - d T^2 + \d_{ij} \, d x^i dx^j  \rs  \rb .			\la{Metric:AdSwithdS}
\ee
When the boundary is located at a fixed $u=\e$, it becomes a $d$-dimensional dS space
\be		 
ds_{dS}^2 = \fr{1}{H^2 T^2} \ls - d T^2 + \d_{ij} \, d x^i dx^j  \rs   ,   \la{Result:BdSmetric}
\ee
where the Hubble constant is given by $H = \e/R^2$. If we further introduce $\ta = i T$ and $u  = i \bar{u}$,  the Lorentzian AdS metric in \eq{Metric:AdSwithdS} becomes a Euclidean one
\be
ds^2 = \fr{R^2}{u^2} \lb \fr{d u^2}{1 - u^2/R^2}  + \fr{R^2}{\ta^2} \ls  d \ta^2 + \d_{ij} \, d x^i dx^j   \rs  \rb ,  \la{Result:EAdSdS}
\ee
which has the same form as that of the Euclidean AdS metric with an AdS boundary in \eq{Result:EAdSwithAdS}. If we evaluate two-point function on this Euclidean background, we can easily derive a Lorentzian correlation function on the boundary dS and AdS space.

\subsection{Two-point function in an expanding universe}

First, we focus on the dual of a QFT living in a dS space and study its two-point functions holographically. When two local operators are located at $\lc u, \ta, \vec{r} \rc=\lc 0, \ta_1, \vec{r}_1 \rc$ and $\lc 0, \ta_2, \vec{r}_2 \rc$, we rearrange the operator's positions to be at $\lc u, \ta, \vec{r} \rc=\lc 0, \ta_1, x_1, 0,\cdots,0 \rc$ and $\lc 0, \ta_2, x_2 , 0,\cdots,0 \rc$ with $|\vec{r}_1 - \vec{r}_2 | = |x_1 - x_2|$ due to the $(d-1)$-dimensional rotational symmetry. For the Euclidean AdS space in \eq{Result:EAdSdS}, a geodesic length connecting two operators is governed by
 \be			\la{Geodesic:HEEAdS3}
L  (T_1, x_1;T_2, x_2) =  \int_{x_1}^{x_2} dx \ \fr{R}{u} \sqrt{\fr{u'^2}{1 - u^2 /R^2} + \fr{R^2}{\ta^2} \ls 1 + \ta'^2  \rs} \ ,
\ee
where $\ta$ and $u$ are considered as functions of $x$ and the prime means a derivative with respect to $x$. It is worth noting that, due to the translation symmetry in the $x$-direction, it is more convenient to take $\ta$ and $u$ as functions of $x$. We first assume that there exists a turning point at $x =  x_t$ where $u'(x_t) = 0$. Due to the translation symmetry in the $x$-direction, there is a conserved charge satisfying 
\be
\frac{\sqrt{R^2- u^2}}{ \ta z \sqrt{\left(1+\ta'^2\right) \left(R^2 - u^2\right)+\ta^2  z'^2}} 
= \frac{1}{ \ta_t z_t \, \sqrt{1 + \ta_t'^2} }  .    	\la{Result:conquan}
\ee
where $u_t = u(x_t)$, $\ta_t = \ta(x_t)$ and $\ta_t' = \ta'(x_t)$ respectively.

Unlike the previous cases, this system has only one conserved charge. Therefore, we have to solve the dynamical equation of $\ta$ and $u$ to determine a geodesic. Note that two dynamical equations in this case are not independent because of the conservation law. Combining two dynamical equations of $u$ and $\ta$, we can find the following decoupled equation of $\ta$ \cite{Chu:2016pea,Koh:2020rti,Park:2020xho}
\be
0 = \ta \ta'' + \ta'^2 + 1 .
\ee
which allows a general solution 
\be
\ta(x) = \sqrt{ c_2^2 - (c_1 - x)^2  }  ,
\ee
where $c_1$ and $c_2$ are two integral constants. Plugging this solution into \eq{Result:conquan}, we obtain
\be
\fr{du}{dx} = \pm \frac{c_2 \sqrt{u_t^2-u^2} \sqrt{R^2-u^2}}{u \left(c_2^2-\left(c_1-x\right){}^2\right)} .
\ee
A solution of this differential equation is given by
\be
u(x) = \frac{\sqrt{R^2 \lb (1+ c_3) x +c_2 - c_1  - (c_1 + c_2 )c_3  \rb{}^2-z_t^2 \lb (1 -c_3) x+c_2 - c_1+  \left(c_1 +c_2 \right) c_3 \rb{}^2}}{2 \sqrt{c_3} \sqrt{c_1-c_2-x} \sqrt{c_1+c_2-x}}  , 
\ee
where $c_3$ is another integral constant. Here, four undetermined parameters, $c_1$, $c_2$, $c_3$, and $z_t$, can be fixed by imposing the following four boundary conditions, $\ta_1 = \ta(x_1)$, $\ta_2 = \ta(x_2)$ and $0= u(x_1)$, and $0= u(x_2)$. First two boundary conditions determine $c_1$ and $c_2$ as
\be
c_1 &=& \fr{\ta_2^2 - \ta_1^2 + x_2^2 - x_1^2}{2 (x_2 - x_1)} , \nn
c_2 &=& - \frac{\sqrt{\left(\tau _1-\tau _2\right){}^2+\left(x_1-x_2\right){}^2} \sqrt{\left(\tau _1+\tau_2\right){}^2+\left(x_1-x_2\right){}^2}}{2 \left(x_1-x_2\right)} .
\ee
The remaining two boundary conditions determine $c_3$ as a function of $c_1$ and $c_2$
\be
c_3 &=&   \frac{ \sqrt{\left(c_1-c_2-x_1\right) \left(c_1-c_2-x_2\right)}}{\sqrt{ \left(c_1+c_2-x_1\right)  \left(c_1+c_2-x_2\right)}} .
\ee
Using these results, we finally determine the turning point $u_t$ in terms of operator's positions
\be
u_t = \frac{R \sqrt{\left(\tau _1-\tau _2 \right){}^2+\left(x_1 - x_2 \right){}^2}}{\sqrt{\left(\tau _1+\tau _2\right){}^2+\left(x_1 - x_2\right){}^2}} .
\ee

The obtained solutions determines the geodesic length as the following form
\be
L  (\ta_1, x_1;\ta_2, x_2) =  \int_\e^{u_t} du \frac{2 R^2 u_t}{z \sqrt{R^2-z^2} \sqrt{u_t^2-z^2}} 
= R  \log \left(\frac{4 R^2 u_t^2}{(R^2-u_t^2) \, \e^2} \right)  ,
\ee
where $\e$ is introduced as a UV cutoff. In the Euclidean dS space, the general two-point function up to normalization reduces to
\be
\bra O \ls \ta_1, \vec{r}_1\rs \ O \ls \ta_2,  \vec{r}_2 \rs \ket   \sim 
 \left(\frac{\tau _1 \tau _2 }{   | \tau _1 - \tau_2  | {}^2+ |\vec{r}_1 - \vec{r}_2 | {}^2 }\right)^{\Delta }.  
 \la{Result:EtwoptdS}
\ee
where $\D$ indicates a conformal dimension of $O$. After the Wick rotation $\ta = i T$, the Lorentzian two-point function becomes
\be
\bra O \ls T_1, \vec{r}_1\rs \ O \ls T_2,  \vec{r}_2 \rs \ket   \sim 
 \left(\frac{  T_1 T_2 }{ - | T_1 -  T_2 |{}^2+ | \vec{r}_1 - \vec{r}_2 |{}^2  }\right)^{\Delta }.  
\ee

In the above calculation, we exploited the conformal time $T$ for convenience. In order to study a two-point function in the expanding universe, it is more convenient to introduce a cosmological time $t$ 
\be
T = \fr{ e^{ - H t} }{H} .
\ee
In terms of the cosmological time, the dS metric in \eq{Result:BdSmetric} is rewritten as
\be		 
ds_{dS}^2 = - d t^2 + e^{2 H t} \, \d_{ij} \, d x^i dx^j    ,
\ee
which describes an eternal inflation. In this eternally expanding universe, the two-point function for $t_2 > t_1$ is given by 
\be
\bra O \ls t_1, \vec{r}_1\rs \ O \ls t_2,  \vec{r}_2 \rs \ket   \sim 
 \frac{ T_1^{2 \D} \ e^{- \D  H \,  | t_2 - t_1 |} }{ | -  T_1^{2} \  (1 - e^{-H  | t_2 - t_1 |} )^2 + (\vec{r}_1 - \vec{r}_2 )^2 |^\D} ,
\ee
where the conformal time is determined as a function of the cosmological time
\be
T_1  =  \fr{e^{-H t_1}}{H} \quad {\rm and} \quad T_2 = T_1 e^{- H  | t_2 - t_1 |} .
\ee

A temporal two-point function in the early time ($ | \vec{r}_1 - \vec{r}_2| \ll  | t_2 - t_1 | \ll1/H $) decreases by a power law
\be
\bra O \ls t_1, \vec{r}_1\rs \ O \ls t_2,  \vec{r}_2 \rs \ket   \approx 
 \frac{ 1}{  | t_1 - t_2 |^{2 \D}} .
\ee
This is the correlator of a CFT. In the late time era ($ | \vec{r}_1 - \vec{r}_2| \ll 1/H  \ll  | t_2 - t_1 | $), the holographic result shows that a temporal two-point function suppresses exponentially
\be
\bra O \ls t_1, \vec{r}_1\rs \ O \ls t_2,  \vec{r}_2 \rs \ket   \approx  e^{- \D  H \,  | t_2 - t_1 |} .
\ee
This is a typical feature of the massive operator's correlator. In this case, $\D H$ plays a role of an effective mass. On the other hand, a spatial two-point function for $ | \vec{r}_1 - \vec{r}_2|  \gg  |t_1 - t_2|$ leads to the following correlator 
\be
\bra O \ls t_1, \vec{r}_1\rs \ O \ls t_2,  \vec{r}_2 \rs \ket   \approx
 \frac{ e^{- \D H  \,  | t_2 - t_1 |} }{ | \vec{r}_1 - \vec{r}_2 |{}^{2 \D} } .
\ee
This shows that the two-point function always suppresses by a power-law in the spatial direction. This implies that the operator behaves as massless one in the spatial direction unlike the temporal correlator. When two operators are located at the same time ($t_1=t_2=t$), a spatial two-point function at the time $t$ is given by
\be
\bra O \ls t, \vec{r}_1\rs \ O \ls t,  \vec{r}_2 \rs \ket   \sim 
 \frac{ T_0^2\,  e^{-2 \D H \, (t - t_0)} }{ | \vec{r}_1 - \vec{r}_2 |{}^{2 \D} },
\ee
where $t_0$ is an appropriate reference time satisfying $T_0 = e^{ - H t_0}/H$. Therefore, the spatial two-point function exponentially suppresses with time due to the expansion of the background spacetime. This is consistent with the results obtained in a different holographic model \cite{Koh:2020rti,Park:2020xho}.

To understand the obtained holographic result further on the dual QFT side, we take into account a QFT living in a $d$-dimensional dS space and discuss its two-point function. A Euclidean metric of a $d$-dimensional dS space can be written as
\be
ds_{dS}^2  = g_{\m\n} dx^\m dx^\n = \fr{R^2}{ \ta^2} \ls d \ta^2 + \d_{ij} dy^i dy^j\rs ,
\ee
where $x^\m = \lc \ta, y^i \rc$ with $i,j = 1, \cdots, (d-1)$ and $\ta$ indicates a Euclidean time. Now, we consider a scalar field on this dS space 
\be
S = \fr{1}{2}  \int d^d x \sqrt{g} \ \ls \pa^\m \ph \,    \pa_\m \ph + \xi \, {\cal R}_{dS}^{(d)} \ph^2  \rs .
\ee  
When the scalar field conformally couples to the dS background, the coefficient $\xi$ is given by
\be
\xi = \fr{(d-2)}{4 (d-1)} .
\ee
In this theory, a conformal dimension of $\phi$ is given by $\D_\phi = (d-2)/2$. Since the background dS space has a positive curvature scalar, ${\cal R}_{dS}^{(d)} = \fr{d (d-1)}{R^2}$, the scalar field in the dS space has an effective mass 
\be
m_{dS}^2 = \fr{(d-2)}{4 (d-1)} {\cal R}_{dS}^{(d)} =  \fr{d(d-2)}{4 R^2}  .   \la{Result:massgap}
\ee
Therefore, the two-point function of $\ph$ satisfies 
\be
\fr{1}{\sqrt{g}} \lb -  \pa_\m \sqrt{g} g^{\m\n} \pa_\n \fr{}{} + m_{dS}^2 \rb  \bra \phi (\ta_1,\vec{y}_1) \ \phi (\ta_2, \vec{y}_2)\ket= \fr{\d ^{(d)}\ls x_1-x _2 \rs }{\sqrt{g}}  .
\ee 
Solving this equation, we can obtain the following Lorentzian two-point function after the Wick rotation ($\ta= i T$)
\be
\bra \phi (\ta_1,\vec{y}_1) \ \phi (\ta_2, \vec{y}_2)\ket  \sim  \ls   \fr{T_1 T_2}{ |-   ( T_1 - T_2 )^2 + ( \vec{y}_1 - \vec{y}_2 )^2| } \rs^{\D_\ph}  .
\ee
 If we further consider an operator, $O = \phi^n$, its two-point function becomes
\be
\bra O (T_1 , \vec{y}_1) \ O (T_2 , \vec{y}_2) \ket  \sim  \ls \fr{T_1 T_2}{| -  (T_1 - T_2 )^2 + ( \vec{y}_1 - \vec{y}_2 )^2 | } \rs^{\D} ,
\ee
where $\D = n \D_\ph$ at the tree level. This is the two-point function obtained in the previous holographic calculation. If we further take into account interactions and their quantum corrections, they can generate an anomalous dimension, which modifies the conformal dimension of the operator. At quantum level, as a result, the conformal dimension of $O$ can be different from the tree-level result \ct{Park:2022mxj}.

\subsection{Two-point correlation in an AdS space}

The previous holographic calculation with the dS boundary can be easily generalized into the case with an AdS boundary because the Euclidean bulk AdS metric in \eq{Result:EAdSwithAdS} is invariant under exchanging time and one of the spatial coordinates. When we express the boundary AdS metric as the following form 
\be
ds^2  = \fr{R^2}{ w^2} \ls d w^2 - d t^2+ \d_{ab} dx^a dx^b \rs ,
\ee
where $w$ ($w \ge0$) corresponds to the radial coordinate and $a,b = 1, \cdots, (d-2)$, the general two-point function in the holographic setup is given by
\be
\bra O \ls t_1, w_1, \vec{x}_1\rs \ O \ls t_2, w_2 , \vec{x}_2 \rs \ket   \sim 
 \left(\frac{w_1 w _2 }{ | -  ( t_1 - t_2 )^2  +  ( w_1 - w_2 )^2+  (\vec{x}_1 - \vec{x}_2 )^2 | }\right)^{\Delta }   .   \la{Result:general2ptAdS}    
\ee
Similar to the dS case, this result shows that the two-point function suppresses by a power-law except for the radial direction $w$. For $t_1 = t_2=t$ and $\vec{x}_1 = \vec{x}_2= \vec{x}$, the two-point function in the radial direction of the AdS space reduces to
\be
\bra O \ls t, w_1, \vec{x} \rs \ O \ls t, w_2 , \vec{x} \rs \ket   \sim  
 \left( \left| \frac{ w_1 + w_2}{ w_1 - w_2} \right|^2 - 1 \right)^{\Delta } .
\ee
When the distance of two operators is short with satisfying $ | w_1 - w_2 | \ll | w_1 + w_2 | $, the two-point function behaves like
\be
\bra O \ls t, w_1, \vec{x} \rs \ O \ls t , w_2 , \vec{x} \rs \ket   \sim \frac{ 1}{ | w_1 - w_2 |^{2 \D} } ,
\ee
which is equivalent to two-point function in the other directions. In the large distance limit satisfying $ | w_1 - w_2 | \approx | w_1 + w_2 | $, however, the two-point function in the $w$-direction shows a different behavior
\be
\bra O \ls t, w_1, \vec{x} \rs \ O \ls t , w_2 , \vec{x} \rs \ket   \sim \frac{ 1}{ | w_1 - w_2 |^{\D} } .
\ee
In other words, the scaling dimension of the operator $O$ changes from $\D$ to $\D/2$ as $| w_1 - w_2|$ increases.

From the QFT point of view, similar to the dS case, the previous holographic two-point function can be understood via a scalar field conformally coupled to the background Euclidean AdS space  
\be
S = \fr{1}{2}  \int d^d x \sqrt{g} \ \ls \pa^\m \ph \,    \pa_\m \ph  + \xi \, {\cal R}_{AdS}^{(d)} \ph^2  \rs.
\ee 
where the scalar field has an effective mass due to the curvature of the background AdS space
\be
m_{AdS}^2 = \fr{(d-2)}{4 (d-1)} {\cal R}_{AdS}^{(d)}  =  - \fr{d(d-2)}{4 R^2} .
\ee 
Therefore, a two-point function of a scalar field must satisfy the following equation in the AdS space 
\be
\fr{1}{\sqrt{g}} \lb -  \pa_\m \sqrt{g} g^{\m\n} \pa_\n \fr{}{} + m_{AdS}^2 \rb  \bra \phi (\ta_1,\vec{y}_1) \ \phi (\ta_2, \vec{y}_2)\ket= \fr{\d ^{(d)}\ls x_1-x _2 \rs }{\sqrt{g}}  .
\ee
The solution of this equation is coincident with the Euclidean version of the holographic result  in \eq{Result:general2ptAdS}.


\section{Discussion}

In this paper, we have studied how to calculate general correlation functions in the holographic setup. After regarding the dual gravity of a QFT, we evaluated the geodesic length connecting boundary operators which is directly related to a boundary-to-boundary green function in the bulk. According to the AdS/CFT correspondence, this Green's function can be regraded as a correlation function of the dual QFT. We showed that this holographic approach reproduces the known correlation functions of CFT, and that it also offers a novel method to understand the scale dependence of correlation functions in various situations like a thermal CFT or QFT in curved spacetimes.

First, we discussed how to calculate general two- and three-point functions of CFT holographically when operators are located at arbitrary positions and times. This is the generalization of the equal-time and equal-position correlation functions studied before \cite{Park:2022abi,Park:2022mxj}. For equal-time and equal-position correlators, there is only one conserved charge in the dual description. For the general correlator, however, there are two conserved charges due to the translational symmetries in the temporal and spatial directions. Exploiting these conserved quantities with appropriate boundary conditions, we determined the exact configuration of the geodesic and evaluated its length analytically. Intriguingly, we showed that this holographic calculation reproduces the exact two-and three-point functions expected by the conformal symmetry.

The holographic approach was also applied to the two-point functions of non-trivial QFTs like a thermal CFT or QFT in an expanding universe. A thermal system or expanding background usually has a parameter, temperature or Hubble constant, specifying the system's scale. Since this parameter is finite, its effect is negligible in the UV limit. Therefore, the correlation function of such a system reduces to the CFT result in the UV limit having a short distance and time interval. In the IR limit, however, the finite correction can give rise to a significant effect on the theory with modifying the correlation functions. Although it is important to understand such IR modification, it is generally hard to calculate the IR correlation function exactly because of the nonperturbative feature of IR physics. In the present work, we calculated the exact correlation functions valid in the entire energy scale by applying the holographic method. The holographic result showed that the correlation functions of a thermal CFT or QFT in expanding universe, as expected, are the same as the CFT results in the UV limit. In other words, the correlation function suppresses by a power-law in the short time interval and distance limit. In the IR limit, however, the correlation functions decreases exponentially due to the screening effect for a thermal CFT or the expansion of the background spacetime for the expanding universe.   

If we further consider relevant operators deforming the UV CFT, they significantly modify the IR physics with providing a nontrivial RG flow. In this procedure, correlation functions also seriously change. Therefore, it would be interesting to investigate the scale-dependent correlators for a system having a nontrivial RG flow. In this case, a UV CFT can flow to a new IR CFT in which the scaling dimension of an operator changes with an anomalous dimension. We hope to report more interesting results on this issue in future work.

\vspace{0.5cm}

{\bf Acknowledgement}

CP acknowledges the hospitality at APCTP where a portion of this work was undertaken. This work was supported by the National Research Foundation of Korea(NRF) grant funded by the Korea government(MSIT) (No. NRF-2019R1A2C1006639). The author J.P. indebted to the authorities of
IIT Roorkee for their unconditional support towards researches in basic sciences.



\end{document}